\documentclass[preprints,article,accept,moreauthors,pdftex]{mdpi}
\firstpage{1} 
\pubvolume{1}
\issuenum{1}
\articlenumber{0}
\pubyear{2022}
\copyrightyear{2022}
\datereceived{} 
\dateaccepted{} 
\datepublished{} 
\hreflink{https://doi.org/} % If needed use \linebreak
\usepackage{bm}

\newcommand{\aap}{Astron.\ Astrophys.}

\newcommand{\prc}{Phys.\ Rev.\ \rm C}
\newcommand{\prl}{Phys.\ Rev.\ Lett.}

\newcommand{\nphysa}{Nucl.\ Phys.\ \rm A}

\Title{Stability of spherical nuclei in the inner crust of neutron stars}

\TitleCitation{Stability of spherical nuclei in the inner crust of neutron stars}

\Author{Nikita A. Zemlyakov $^{1,2}$* and Andrey I. Chugunov $^{1}$\orcidA{}}
\AuthorNames{Nikita A. Zemlyakov and Andrey I. Chugunov}

\AuthorCitation{Zemlyakov, N.A.; Chugunov, A.I.}
\address{%
$^{1}$ \quad Ioffe Institute, 26 Politekhnicheskaya st., St. Petersburg 194021, Russia; \\
$^{2}$ \quad Peter the Great St.\ Petersburg Polytechnic University, 29 Politekhnicheskaya st., St.\ Petersburg 195251, Russia;}

\corres{Correspondence: zemnic5@gmail.com}

\abstract{Neutron stars are the densest objects in the Universe. 
In this paper we consider so-called inner crust -- the layer, where neutron-excess nuclei
are
immersed into degenerate gas of electrons and  sea of quasi-free neutrons. 
It was generally believed that spherical nuclei
become unstable with respect to  quadrupole deformations at high densities and here we consider this instability.  Within perturbative approach we show that spherical nuclei with equilibrium number density are, in fact, stable with respect to 
infinitesimal quadrupole deformation.
This is due to  background of degenerate electrons and associated electrostatic potential which maintain stability of spherical nuclei. 
However,  if the number of atomic nuclei per unit volume is much less than the equilibrium value, instability can arise.
To avoid confusion we stress that our results are limited to infinitesimal deformations and do not guaranty strict thermodynamic stability  of spherical nuclei. In particular, they does not exclude that substantially non-spherical nuclei (so-called pasta phase) represent thermodynamic equilibrium state of the densest layers of neutron star crust. Rather our results points that spherical nuclei can be metastable even if they are not energetically favourable and the timescale of transformation of spherical nuclei to the pasta phases should be estimated subsequently.
}


\begin{document}

\section{Introduction}

The inner crust of neutron stars extends from the density $\rho_\mathrm{drip} \approx 4.3 \times 10^{11}$~g/cm$^3$ to $\sim 10^{14}$~g/cm$^3$  \cite{HPY07_book,CH08_NScrust}. It consists of fully ionised atomic nuclei immersed into background of quasi-free neutrons and relativistic degenerate electron gas. Atomic nuclei have large neutron excess because of the high chemical potential of the electrons. They have a spherical shape in most of the inner crust, but in the deepest layers energetically favourable nuclei configuration can become substantially non-spherical  (cylinders, planes and inverse configurations; so-called pasta phases)  \cite{RPW83_pasta,Hashimoto_ea84,HPY07_book,CH08_NScrust,Newton_ea13_Survey,CH17}
(see also \cite{DinhThi_ea21,Pearson_ea20_Pasta,Newton_ea21_Glassy_pasta,pc22} for the most recent progress).% 
\footnote{Some works  (e.g., \cite{DH00_InnerEdge,Vinas_ea17}) predict pasta phases to be absent in neutron stars.}

\begin{figure}
	\includegraphics[width=16.4 cm]{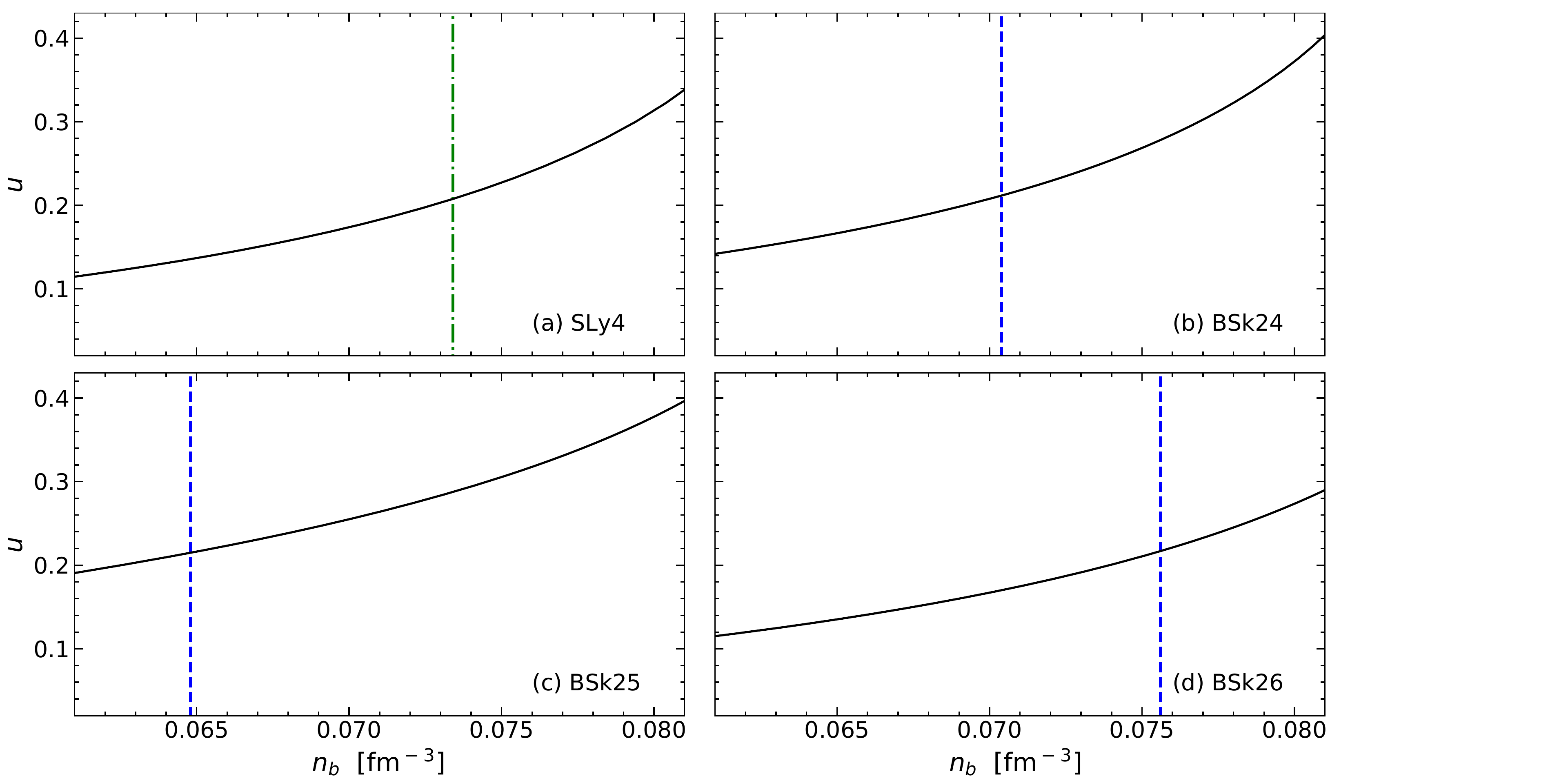}
	\caption{The filling factor for spherical nuclei as a function of nucleon number density. Panels (a), (b), (c), (d) are plotted for the Skyrme potentials SLy4 \cite{Chabanat_ea97_SLY4,Chabanat_ea97_SLY4_p2_nuclei} and BSk24, BSk25, BSk26 \cite{Goriely_ea13_BSK24}, respectively. The black solid line is the filling factor, the  vertical lines represent maximal density there spherical nuclei predicted to be energetically favourable. For SLy4 model (panel a, dash-dot line) it corresponds to the transition from spherical to the uniform nuclear matter (neutron star core), while for BSk models (panels b-d; dashed line)  transition  to the cylindrical nuclei takes place.\label{fig1}
		The plot is based on calculations of Ref.\ \cite{Zemlyakov_2021}.}
\end{figure}

Following \cite{pr95}, the density region for spherical nuclei was generally assumed to be limited by instability of spherical nuclei  with respect to quadrupole deformations (see, e.g., \cite{iws02,HPY07_book,CH08_NScrust}).
Namely, applying the instability criterion  derived by Bohr and Wheeler \cite{BW39} for isolated nucleus, the spherical nuclei were predicted to become 
absolutely unstable
(even for infinitesimal deformation) when the ratio of the nucleus volume to the Wigner-Seitz cell volume (filling factor) $u$ reaches a value of $1/8=0.125$ \cite{pr95}. 
Indeed, recent extended Thomas-Fermi calculations of Ref.\ \cite{Pearson_ea20_Pasta} reports a transition from spherical to cylindrical nuclei at filling factor close to $0.125$ (see their table  XII and respective discussion).
However, 
calculations   within the compressible liquid drop model (CLDM) typically predict spherical nuclei to be energetically favourable up to the larger filling factor $\sim 0.2$ (see, e.g., \cite{Hashimoto_ea84,DH00_InnerEdge} and Fig.~\ref{fig1} for numerical illustration). 
Thus, at least within CLDM approach, there is a contradiction  between numerical results and predicted instability: spherical nuclei are predicted to be thermodynamically stable at the region there they supposed to be absolutely unstable. Obviously, it have two (not mutually exclusive) solutions: A)  the true thermodynamic equilibrium for filling factors 0.125-0.2 correspond to complex nuclear structures (e.g., \cite{Newton_ea21_Glassy_pasta}), which stays beyond the scope of 
most works based on CLDM approach; B) the instability  of spherical nuclei is suppressed in the inner crust.

%%%%%%%%%%%%%%%%%%%%%%%%%%%%%%%%%%%%%%%%%%

The option B), enhanced stability of spherical nuclei, were suggested by a number of authors (e.g., \cite{DH00_InnerEdge,Watanabe10}). The strongest point in support of this solution was suggested in  Ref.\ \cite{KW21}, which, in particular, investigated the stability of spherical nuclei
with respect to infinitesimal deformations within the Wigner-Seitz cell approximation.%
\footnote{According to \cite{pr95}, the corrections to  Bohr and Wheeler \cite{BW39} instability  conditions associated with presence of other nuclei and electrons were already calculated by Brandt \cite{brandt1985kernestof}. However, the latter work is unavailable for us. Note, that authors of  \cite{pr95} neglect these corrections.
}
Authors concluded that spherical nuclei are stable at arbitrary filling factor, if  the number density of nuclei corresponds to the optimal value.%
\footnote{So-called virial theorem,  derived by \cite{Baym_ea71} and applied by authors of Ref.\ \cite{KW21}  (see Eq.\ (\ref{equilibrium}) here), 
is applicable only for optimal number density of nuclei.}
However, in our opinion, results of Ref.\ \cite{KW21} are based on inaccurate boundary conditions. Namely, to ensure that the electric field flux over the cell boundary is zero, authors of Ref.\ \cite{KW21} impose the Nuemann boundary condition, i.e. demand that  the normal component of the electric field is zero at each point of the cell boundary.  The latter seems unreasonable: thanks to the zero charge of the cell and Gauss theorem,  the electric field flux over the cell boundary vanishes automatically if electron and nucleus (proton) contributions are both correctly incorporated.
Imposing of specific boundary conditions, in particular, the Neumann boundary condition,  is equal to assumption that the charges outside the cell rearrange themselves  to a distribution  which is required to ensure imposed boundary condition. Note, that the required distribution of outside charges depends on deformation of nucleus, making such rearranging unnatural from our point of view.
Here we correct this problem. In addition, our consideration naturally takes into account the neutron skin, which is essential  ingredient of two phase system boundary thermodynamics (e.g., \cite{ll80,lpr85}),  however, account of this effect do not change the results.

To avoid confusion, let us stress that we analyse only stability with respect to infinitesimal deformations. Clearly, it does not sufficient to guaranty absolute thermodynamic stability.
Indeed, as shown by Hashimoto et al.\ \cite{Hashimoto_ea84}, the spherical nuclei can not be energetically favourable at too large filling factors and thus spherical nuclei can be treated only as metastable
in strict thermodynamic sense. 

\section{Calculations}
In order to explain the nature of the above mentioned contradictions we checked validity of the option B): suppression of spherical nuclei instability  in inner crust. We applied  CLDM of Ref.\ \cite{GC20}, in which the nucleus is surrounded by quasi-free neutrons, being located at the center of a spherical Wigner-Seitz  cell, which is uniformly filled with electrons (according to the quasi-neutrality requirement the total charge of the cell is zero). This model naturally incorporates neutron skin effects (see supplementary material to Ref.\ \cite{GC20} for details).

Following Bohr and Wheeler \cite{BW39}, we considered the change of energy when the nucleus is deformed from a sphere to an
spheroid with semi-axes $R\left(1+\varepsilon \right)$ and $R/\sqrt{1+\varepsilon}$. Here $R$ is the radius of the spherical nucleus, $\varepsilon$ is infinitesimal deformation parameter. The Wigner-Seitz cell assumed to stay spherical. 
Within CLDM of Ref.\  \cite{GC20} the change of the cell energy can be calculated analytically  up to the second order in $\varepsilon$ (see Appendix for the details):
\begin{equation}
\delta E= \left[\frac{8\pi R^2}{5}\sigma + \frac{3}{5} \left(\frac{u}{2}-\frac{1}{5}\right) \frac{Z^2e^2}{R} \right]\varepsilon^2.
\label{energy_change}
\end{equation}
Here $\sigma$ is the surface tension,
$R$ is the 
radius of the nucleus before deformation,
$Ze$ is the nucleus charge, $e$ is the elementary charge.
Eq.\ (\ref{energy_change}) is similar to Eq.\ (9) of Ref.\ \cite{BW39}, but does not coincide exactly (even for $\varepsilon^2$ terms, considered here) due to presence of electron background, which induces electrostatic potential and modifies the Coulomb energy change associated with nuclei deformation.
The difference is $\propto u$.
It agrees with results of Ref.\ \cite{brandt1985kernestof}, as they were cited in \cite{pr95}. However, as we demonstrate below, this  term  can not be neglected  because it leads to suppression of the instability.

It is worth to stress that Eq. (\ref{energy_change}) holds true in exactly the same form for a simplified CLDM, which neglects neutron skin effects (see Appendix for derivation details), so  our results are equally applicable for this widely applied type of CLDMs.

\subsection{Equilibrium inner crust}
The virial theorem \
\cite{Baym_ea71}
 (see also supplementary material in \cite{GC20} for derivation with accurate account of neutron adsorption effects)
provide coupling for Coulomb energy and surface terms, if crust composition corresponds to the equilibrium (the number of atomic nuclei per unit volume is optimal)  :
\begin{equation}
4\pi \sigma R^2=2 \frac{3}{5}\frac{Z^2e^2}{R} \left(1-\frac{3}{2}u^{\frac{1}{3}}+\frac{1}{2}u \right).
\label{equilibrium}
\end{equation}

Substituting $4\pi \sigma R^2$, given by  (\ref{equilibrium}), into Eq.\ (\ref{energy_change}) we got the energy change associated with infinitesimal nuclei deformation in equilibrium crust:
\begin{equation}
\delta E =\frac{3}{5}\frac{Z^2e^2}{R} \left(\frac{3}{5}-\frac{6}{5}u^{\frac{1}{3}}+\frac{9}{10}u \right)\varepsilon^2.
\label{change_en_equil}
\end{equation}
It is easy to check that at any value of the filling factor $u \in (0,1)$ the energy change $\delta E$ remains positive. Thus, the nuclei in the crust with equilibrium composition remain stable with respect to infinitesimal  quadrupole deformations at any values of the filling factor. 
It is worth to stress that this result is derived analytically and stays the same for CLDMs which includes and neglects neutron skin effects. In particular, it does not depend on the choice of nuclear physical model required to specify numerical parameters of CLDM (e.g. the bulk energy and surface tension).
Note that the equilibrium composition is believed to be a good model for nonaccreting neutron stars \cite{HPY07_book} (see however \cite{cfg20,pc21}).

\subsection{Non-equilibrium inner crust}
In the more general case the number of atomic nuclei per unit volume 
can differ from equilibrium value.
In this case, instead of (\ref{equilibrium}) we can write  more general expression \cite{GC20}:
\begin{linenomath}
\begin{equation}
4\pi \sigma R^2-2 \frac{3}{5}\frac{Z^2e^2}{R} \left(1-\frac{3}{2}u^{\frac{1}{3}}+\frac{1}{2}u \right)=3\mu_N.
\label{non-equilibrium}
\end{equation}
\end{linenomath}
Here $\mu_N$ is the the chemical potential of the nucleus, which describes the change in energy when one nucleus is created from nucleons, which are already present in the substance (for equilibrium crust $\mu_N=0$). 

Combining equations (\ref{energy_change}) and (\ref{non-equilibrium}), we obtain  the energy change associated with deformation of nucleus in non-equilibrium crust 
\begin{linenomath}
\begin{equation}
\delta E =\frac{3}{5}\frac{Z^2e^2}{R} \left(\frac{3}{5}-\frac{6}{5}u^{\frac{1}{3}}+\frac{9}{10}u \right)\varepsilon^2 + \frac{6}{5}\mu_N \varepsilon^2.
\label{change_en_nonequil}
\end{equation}
\end{linenomath}

For accreting neutron stars, there is an excess of nuclei in the crust \cite{GC20}. It leads to $\mu_N > 0$ and makes the spherical nuclei even more stable  with respect to quadrupole deformations.%
\footnote{To avoid confusion, let us point that the instability considered in \cite{GC20} is not associated with deformation of nuclei and this study does not alter any results of \cite{GC20}.
}
However, if $\mu_N < 0$, i.e., when the number density of nuclei in the stellar matter is less than the equilibrium value, the nuclei may become unstable with respect to quadrupole deformations and this instability likely leads to fission. 

\begin{figure}[H]
\includegraphics[width=\textwidth]{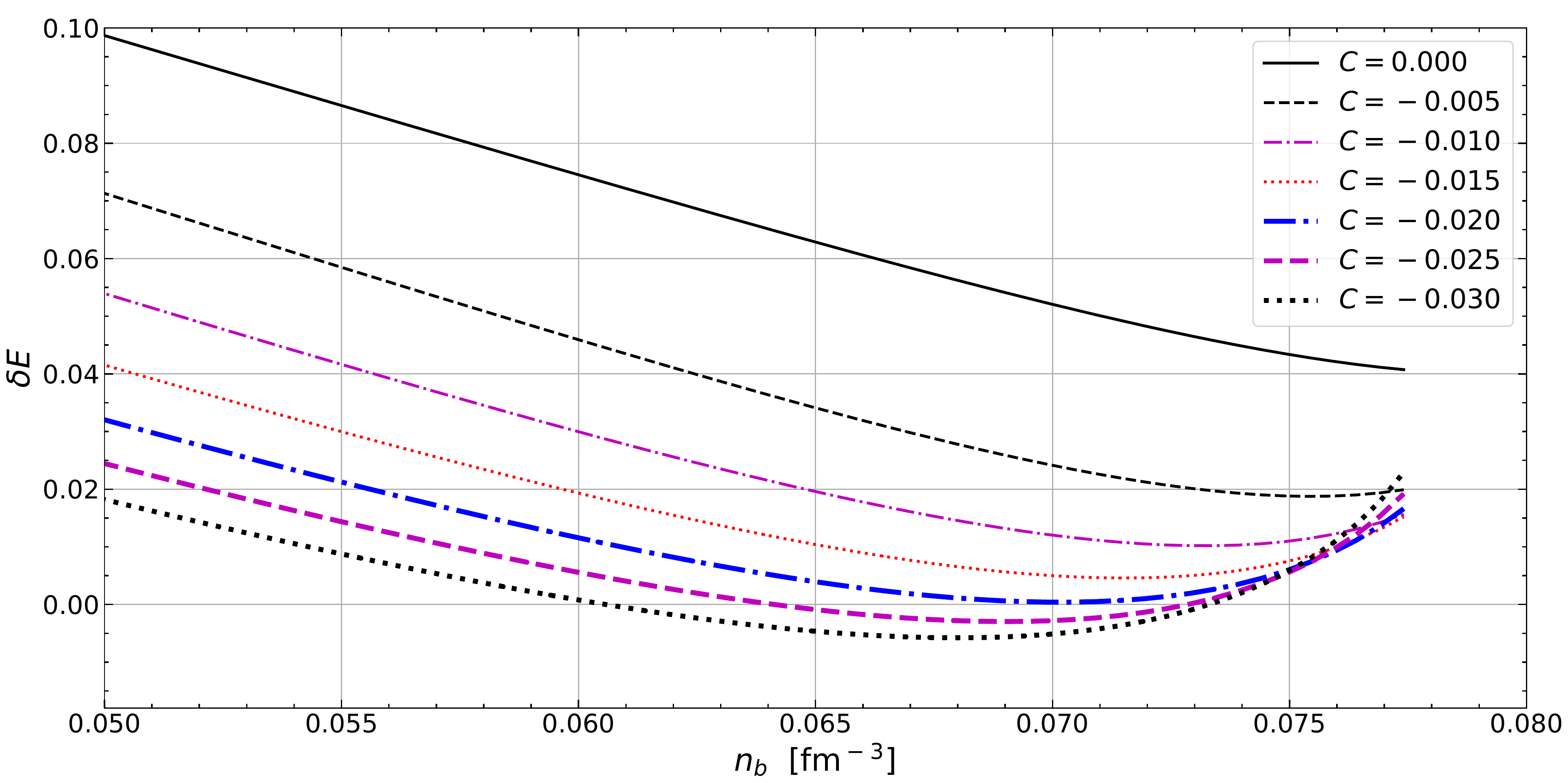}
\caption{$\delta E$ normalized to  $Z^2 e^2\varepsilon^2/R$ as a function of nucleon number density for different values of $C=\mu_N /\mu_n$.  The neutron skin effects are included within CLDM of Ref.\ \cite{GC20}. SLy4   potential \cite{Chabanat_ea97_SLY4,Chabanat_ea97_SLY4_p2_nuclei} is applied. \label{fig2}}
\end{figure}

For numerical illustration of possible instability we  perform calculations including neutron skin (adsorption) effects within CLDM of Ref.\ \cite{GC20} and apply  SLy4 nucleon-nucleon potential \cite{Chabanat_ea97_SLY4,Chabanat_ea97_SLY4_p2_nuclei}. We parametrize matter by baryon number density $n_{b}$ and parameter $C=\mu_N /\mu_n$, which stays constant in the inner crust thanks to diffusion/hydrostatic equilibrium of quasi-free neutrons \cite{GC20}. 
The results are shown in Fig.\ \ref{fig2} for several values of $C$.

%%%%%%%%%%%%%%%%%%%%%%%%%%%%%%%%%%%%%%%%%%
\section{Discussion, results and conclusions}
Within CLDM we demonstrate that spherical nuclei are stable with respect to 
infinitesimal quadrupole deformations, if their number density correspond to the equilibrium value. 
The suppression  of the instability, in comparison with isolated nuclei, is due to the fact that nuclei in the inner crust are immersed into  background of degenerate electrons. The electron charge density is comparable with the charge density of the nucleus and
induces electrostatic potential, which  supports  spherical shape of nuclei.

For non-equilibrium crust, when the number of nuclei per unit volume is less than the equilibrium number, the instability with respect to  infinitesimal quadrupole deformations can appear. This results seems to be natural: fission, likely caused by instability, leads to an increase in the nuclei number, driving composition closer to the equilibrium. According to our calculations for SLy4 potential \cite{Chabanat_ea97_SLY4,Chabanat_ea97_SLY4_p2_nuclei}, nuclei number density should be lower than the equilibrium value for a factor of 2.2-2.4 to ensure the instability at $n_b > 0.047$~fm$^{-3}$.
If nuclei number density is larger than in the equilibrium at the same baryon density  (e.g., in the accreted crust \cite{GC20}), the spherical nuclei are stable with respect to  infinitesimal deformation.
However, it does not exclude other types of instability which are not associated with deformation of nuclei, for example, the instability considered in \cite{GC20}.

Qualitatively similar results were obtained in \cite{KW21}, but as we point in introduction, this work appeal to artificial boundary conditions. It leads to quantitative differences with our results. 

It is worth to stress that our analysis is perturbative, and thus limited to infinitesimal deformations. Obviously, it can not guaranty absolute thermodynamic stability of spherical nuclei. And indeed,  the cylindrical and other pasta phases shown to become more energetically favourable at large filling factors, $u\gtrsim 0.2$ \cite{Hashimoto_ea84}.
Combination of stability of spherical nuclei for infinitesimal deformations with instability in strict thermodynamic sense for $u\gtrsim 0.2$ suggest that in this conditions spherical nuclei should be, in fact, metastable (i.e. correspond to local energy minimum, which differs from the global minimum). It opens an interesting task to estimate the transition timescale from spherical to nonspherical shapes, however, we leave any estimates for this timescale beyond of this work.
However, we point that $u=1/8$ criterion should not be applied to put an upper bound for the density region of spherical nuclei in equilibrium crust because it was suggested on the base of simplified consideration of fission instability, which is not supported by more detailed analysis, which was performed in our work.

Concluding, we should warn the reader than all our results are obtained within spherical Wigner-Seitz cell approximation, which  is likely inaccurate and should be elaborated for very large filling factor $u\sim 1$.
However, in realistic models of crust the filling factor for spherical nuclei is $u\lesssim 0.2$ and spherical Wigner-Seitz cell seems a reasonable approximation, but we don't check the latter statement straightforwardly. 
We also warn that we neglect the curvature corrections to the surface tension, which shown to be important for thermodynamically determined boundaries of the pasta layers (e.g., \cite{DinhThi_ea21_EPJA}), however, we don't expect that it can affect our results qualitatively.

\vspace{6pt}

%%%%%%%%%%%%%%%%%%%%%%%%%%%%%%%%%%%%%%%%%%
\authorcontributions{Derivations, figures and original draft preparation N.A.Z; formulation of the problem, final edition of the draft and supervision A.I.C. All authors have read and agreed to the published version of the manuscript.}

\funding{Work of N.A.Z. was funded was funded by the Russian Science Foundation grant number 19-12-00133-P.
	\url{https://rscf.ru/project/19-12-00133/} {(accessed on June 30 2022). }
}

\dataavailability{Not applicable.}

\conflictsofinterest{The authors declare no conflict of interest. The funders had no role in the design of the study; in the collection, analyses, or interpretation of data; in the writing of the manuscript, or in the decision to publish the~results.} 

%%%%%%%%%%%%%%%%%%%%%%%%%%%%%%%%%%%%%%%%%%
%% Optional
\abbreviations{Abbreviations}{
The following abbreviations are used in this manuscript:\\

\noindent 
\begin{tabular}{@{}ll}
CLDM & Compressible liquid drop model\\
SLy & Skyrme-Lyon\\
BSk & Brussels-Skyrme
\end{tabular}}

\appendixtitles{yes} % Leave argument "no" if all appendix headings stay EMPTY (then no dot is printed after "Appendix A"). If the appendix sections contain a heading then change the argument to "yes".
\appendixstart
\appendix

\section{Derivation of the energy change associated with nuclei deformation (Eq.\ \ref{energy_change})}
At the appendix we derive Eq.\ (\ref{energy_change}) which describes the change of the cell energy associated with deformation of nuclei.
The derivations are provided for two versions of CLDM, one which neglects neutron skin effects and model of Ref.\ \cite{GC20}, which incorporates these effects thermodynamically consistently. 
The curvature corrections are neglected in both cases.
In the first subsection \ref{Sec_CoulEnergChange} we derive the Coulomb energy of a cell with deformed nuclei, which is required for derivation of Eq.\ (\ref{energy_change}).

\subsection[\appendixname~\thesection]{Coulomb energy of the cell with deformed nucleus}
\label{Sec_CoulEnergChange}

We start from consideration of the Coulomb energy of a cell with deformed nucleus in the center.
It is essentially electrostatic problem: calculation of the Coulomb energy of uniformly negatively  charged sphere with radius $R_{c}=R/u^{1/3}$ and uniformly positively charged spheroid (with semi-axes $R\left(1+\varepsilon \right)$ and $R/\sqrt{1+\varepsilon}$), located at the center of the sphere. The net charge of this system is zero. We neglect terms $\propto \varepsilon ^3$.

We begin with the Poisson's equation for Wigner-Seitz cell
\begin{equation}
\Delta \varphi = -4 \pi \left[\rho_{p} \Theta(R_{sp}(\theta)-r)+\rho_{e}\Theta \left(\frac{R}{u^{1/3}}-r\right)\right],
\label{Poisson}
\end{equation}
where 
where $r$ and $\theta$ are radial distance and the polar angle of spherical coordinate system ($r=0$ is the center of the cell),
$\Theta(x)$ is the Heaviside step function.
$\rho_{pi}=Ze/(4\pi R^3/3)$  is the proton charge density inside nucleus, 
$\rho_{e}=-u \rho_{pi}$ is the electron charge density (the cell is electrically neutral).
The protons are located within a spheroid, with boundary given by
\begin{equation}
R_{sp}(\theta) = \frac{R}{\sqrt{(1-cos^2 \theta)(1+\varepsilon) +cos^2 \theta/(1+\varepsilon)^2 }}.
\end{equation}
It is worth to stress, that $\rho_{p}$ does not depend on $\varepsilon$ because we consider deformation which do not affect nucleus volume.

The solution to Eq. (\ref{Poisson}) can be presented as a sum of the proton potential $\varphi_{p}$ and the electron potential $\varphi_{e}$, while the total Coulomb energy can be presented as $E_C = E_C^{p-p}+E_C^{e-p}+E_C^{e-e}$,
where terms $E_C^{p-p}$, $E_C^{e-p}$, and $E_C^{e-e}$ are proton-proton, electron-proton, and electron-electron 
contributions, respectively:
\begin{eqnarray}
E_C^{p-p}&=&
\frac{1}{2}\int\rho_{p} (\bm{r})\varphi_{p}(\bm r)  d^3 \bm {r} 
=\frac{1}{2}\int_{r<R_{sp}(\theta)}\rho_{p} \varphi_{p}(\bm r)  d^3 \bm {r},
\label{Epp}\\
E_C^{e-p}&=&
\int\rho_{p}(\bm{r}) \varphi_{e}(\bm{r})  d^3 \bm {r}
=
\int_{r<R_{sp}(\theta)}\rho_{p} \varphi_{e}(\bm r)  d^3 \bm {r},
\label{Eep}\\
E_{C}^{e-e}&=&\frac{1}{2}\int\rho_{e}(\bm r) \varphi_{e}(\bm r) d^3 \bm {r}
=\frac{1}{2}\int_{r<R_{c}}\rho_{e}\varphi_{e}(\bm r) d^3 \bm {r}.
\label{Eee}
\end{eqnarray} 
Here we take into account that proton density is zero outside the spheroid.

The explicit form of electron potential inside cell is well known
\begin{equation}
\varphi_{e}=\frac{2 \pi \rho_{e} R^2} {u^{2/3}}-\frac{2}{3}\pi \rho_{e} r^2,
\end{equation}
while the proton potential inside the nucleus ($r<R_{sp}(\theta)$) can be written in form:
\begin{equation}
\varphi_{p}=2 \pi \rho_{p} R^2-\frac{2}{3}\pi \rho_{p} r^2 + \frac{4}{5}\pi \rho_{p} r^2 \varepsilon \left(\frac{3}{2}cos^2 \theta -\frac{1}{2} \right) - \frac{2}{5}\pi \rho_{p} R^2 \varepsilon^2 +\ldots,
\end{equation}
The omitted  terms contributes to the Coulomb energy only at order $\varepsilon^3$ or higher, which is not considered here.
The proton potential outside nucleus is not required to calculate Coulomb energy  (see Eqs.\ \ref{Epp}-\ref{Eee}) and thus not shown here. 

The integrals in Eqs.\ (\ref{Epp})-(\ref{Eee}) can be calculated analytically. Up to the second order in $\varepsilon$ they are:
\begin{eqnarray}
E_ C^ {p-p}&=&\frac{3}{5} \frac{(Ze )^2}{R} -\frac{3}{25} \frac{(Ze )^2}{R} \varepsilon^2,\\
E_ C^ {e-p}&=&-\frac{3}{2} \frac{(Ze )^2}{R}u^{1/3}+\frac{3}{10} \frac{(Ze )^2}{R}u \left(1+\varepsilon^2 \right),\\
E_ C^ {e-e}&=&\frac{3}{5} \frac{(Ze )^2}{R} u^{1/3}.
\end{eqnarray}
For $\varepsilon=0$ they agree with well known expression for the Coulomb energy of the cell with spherical nucleus (e.g., \cite{Baym_ea71}).

\subsection{Calculation of the Energy change neglecting neutron skin}
Within CLDM, which neglects neutron skin the surface tension is typically assumed to be function of $x_i$ -- the ratio of proton number density to the total baryon number density inside the nucleus.
To derive Eq.\ (\ref{energy_change}) we consider difference of the cell energies between two configurations: (a) WS cell with spherical nucleus and (b) WS cell with deformed nucleus.
We assume that the proton and neutron number densities inside nuclei as well as neutron number density outside nucleus are the same in both configurations. The volume of nucleus is also unmodified by deformation.
In this case the energy change associated with deformation, $\delta E=E_b-E_a$, contain only surface and Coulomb terms, while the bulk terms are cancel out.

The surface tension is the same in both configurations ($x_i$ is not  modified) and the surface energy change is given by difference between the surface areas of the spheroid and the sphere. Thus, change of the surface energy is
\begin{equation}
\delta E_ s=\frac{8}{5} \pi \sigma R^2\varepsilon^2.
\label{surface}
\end{equation}

The change of the Coulomb energy can be calculated using results of section \ref{Sec_CoulEnergChange}, which leads to
\begin{equation}
\delta E_ C=\frac{3}{5}\frac{(Ze )^2}{R} \left(\frac{u}{2}-\frac{1}{5} \right)\varepsilon^2.
\label{Coulomb}
\end{equation}

The net energy change $\delta E=\delta E_ C+\delta E_ s$, thus summing up Eq. (\ref{Coulomb}) and (\ref{surface}), we obtain Eq.\ (\ref{energy_change}).

\subsection[\appendixname~\thesection]{Calculation of the energy change including neutron skin}

Derivation of Eq.\ (\ref{energy_change}) for CLDM of Ref.\ \cite{GC20}, which accounts for neutron skin (adsorption) effects, is more complicated. It is due to the fact that nuclei deformation changes of nuclei surface area  and thus amount of neutrons adsorbed on it. Thus, one can not assume that neutron number densities and nucleus volume are not modified by deformation because it will lead to variation of total number of neutrons (and thus net baryon density).

To derive  Eq.\ (\ref{energy_change}) we consider an extended version CLDM of Ref.\ \cite{GC20}, which is allowed for deformed nuclei. Namely, nucleus deformation parameter $\varepsilon$ is added as additional independent CLDM variable. It allows to write down the expression for the energy density $\epsilon$, which differs from Eq.\ (2) of the supplementary material to Ref.\ \cite{GC20} only by the Coulomb and surface energy terms:
\begin{equation}
\epsilon=u\,\epsilon^\mathrm{bulk}(n_{ni},n_{pi})
+(1-u)\,\epsilon^\mathrm{bulk}(n_{no},0)
+\frac{E_{s}(\nu_ s, R,\varepsilon)}{V_{c}}
+\frac{E_{C}(n_{pi}, R,u,\varepsilon)}{V_{c}}
+e_{e}(n_{e}).
\label{epsilon}
\end{equation}
Here $\epsilon^\mathrm{bulk}(n_{n},n_{p})$ is the energy density of bulk nuclear matter at respective neutron and proton number density ($n_{n}$ and $n_{p}$),
$n_{ni}$, $n_{pi}$ are the neutron and proton number density inside nucleus, 
$n_{no}$ is the neutron number density outside nucleus (we assume that proton drip does not take place), and $e_{e}(n_{e})$ is the energy density of degenerate electrons at electron number density $n_{e}=u\,n_{pi}$.
The cell volume $V_{c}=4\pi R^3/(3u)$.
According to the results of section \ref{Sec_CoulEnergChange},
the Coulomb energy is
\begin{equation}
E_{C}=\frac{3}{5}\frac{Z^2e^2}{R} \left[f(u)+\left(\frac{u}{2}-\frac{1}{5}\right)\varepsilon^2\right]
=\frac{16 \pi^2}{15} (n_{pi} e)^2 R^5 \left[f(u)+\left(\frac{u}{2}-\frac{1}{5}\right)\varepsilon^2\right],
\end{equation} 
where $f(u)=1-3\,u^{1/3}/2+u/2$.
 The surface energy
\begin{equation}
E_{s}=\mathcal A \left(\mu_{ns}\nu_ s+\sigma\right),
\end{equation} 
where $\mu_{ns}$ and $\nu_ s$ are chemical potential and surface density of the adsorbed neutrons.
The nuclei surface area $\mathcal A$ is given by the surface area of the spheroid
\begin{equation}
\mathcal A =4\pi R^2+\frac{8}{5}\pi R^2 \varepsilon^2.
\end{equation} 
The thermodynamic consistency requires
\begin{equation}
\frac{d \sigma}{d \nu_{s}}=-\nu_s \frac{d \mu_{ns}}{d \nu_s}.
\label{ThermCons_sigma}
\end{equation}
Thus $\sigma$ and $\mu_{ns}$ can be treated as functions of $\nu_{s}$.

Following \cite{GC20} let us introduce auxiliary variables, which simplifies subsequent analysis
\begin{eqnarray}
n_{n{ i}}^{\rm (tot)} &\equiv& n_{n{i}} u,
\label{nnitot}\\
n_{p{i}}^{\rm (tot)} &\equiv& n_{p{ i}} u,
\label{npitot}\\
n_{n{ o}}^{\rm (tot)} &\equiv& n_{n{ o}} (1-u),
\label{nnotot}\\
n_{ns}^{\rm (tot)}&\equiv& \frac{N_{ s}}{V_{\rm c}},
\label{nstot}
\\
n_N&=&V_{\rm c}^{-1}.
\end{eqnarray}
Here $N_{s}=\mathcal A \nu_ s$ is the number of neutrons, adsorbed to the nucleus.

Here we analyse stability with respect to infinitesimal deformation of nuclei, described by parameter $\varepsilon$.
To do so we check for two conditions: (a) $\varepsilon=0$ is an extremum of the energy density and
(b) the extremum at $\varepsilon=0$ is a (local) minimum.
While doing this analysis we assume that net baryon number density 
\begin{equation}
n_{b}=n_{n{ i}}^{\rm (tot)}+n_{p{ i}}^{\rm (tot)}+n_{n{ o}}^{\rm (tot)}+n_{ns}^{\rm (tot)}
\label{n_b}
\end{equation} 
as well as nuclei number density $n_N$ are constants, while the remaining (internal) CLDM parameters
$n_{p{i}}^{\rm (tot)}$, $n_{o{i}}^{\rm (tot)}$, $n_{ns}^{\rm (tot)}$, and $u$, generally, can vary with variation of $\varepsilon$.

As the first step, we need to specify equlibrium values of internal CLDM parameters, which are given by the condition that the partial derivatives with respect to each of internal parameter is zero.
As long as we are interested on these parameters at $\varepsilon=0$, i.e. for spherical nuclei,  respective equations are exactly the same as in \cite{GC20}.
In particular, if $n_N$ differs from the equilibrium value, Eq.\ (\ref{non-equilibrium}) holds true.

Let us note that the energy density explicitly depends only on $\varepsilon ^2$, thus the partial derivative $\partial \epsilon/\partial \varepsilon\propto \varepsilon$, thus $\varepsilon=0$ is indeed an extremum.

To check is the extremum at $\varepsilon=0$ minimum or maximum we, firstly, note that the mixed partial derivatives 
$\partial^2 \epsilon/\partial \varepsilon\partial p\propto \varepsilon$, and thus they are zero, if calculated at $\varepsilon=0$ (here $p$ is arbitrary parameter of the CLDM model, except $\varepsilon$).
As a result, the variation of energy density associated with infinitesimal $\varepsilon$ is
\begin{equation}
\delta \epsilon=\frac{1}{2}\left.\frac{\partial ^2 \epsilon}{\partial \varepsilon ^2}\right|_{\varepsilon=0}\varepsilon^2.
\end{equation} 
The bulk terms in the energy density do not depend on $\varepsilon$ explicitly and only Coulomb and surface energy contributes to $\partial ^2 \epsilon/\partial \varepsilon ^2$.
These derivatives can be easily calculated,%
\footnote{As long as $\epsilon$ depends on $\varepsilon$ only via $\varepsilon^2$, 
$\left. \partial ^2 \epsilon/\partial \varepsilon ^2\right|_{\varepsilon=0}=
2\left. \partial \epsilon/\partial \left(\varepsilon ^2\right)\right|_{\varepsilon^2=0}$.	
While calculating derivative of the surface energy, one should take in mind that it is calculated at fixed $n_{ns}^{\rm (tot)}$ and thus fixed total amount of adsorbed neutrons. In this case $\delta E_s=\sigma \delta \mathcal A$ due to Eq.\ \ref{ThermCons_sigma}.}
leading to
\begin{equation}
\delta \epsilon=n_N\,\left[\frac{8\pi}{5}  R^2\sigma+\frac{3}{5}\frac{Z^2e^2}{R}\left(\frac{u}{2}-\frac{1}{5}\right)\right]
\varepsilon^2.
\label{delta_epsilon}
\end{equation} 
Multiplying   $\delta \epsilon$ to the cell volume $V_c$ we arrive to Eq.\ (\ref{energy_change}), which describes change of the cell energy associated with nucleus deformation. Positiveness of this energy change guaranty stability with respect to infinitesimal deformations.

%%%%%%%%%%%%%%%%%%%%%%%%%%%%%%%%%%%%%%%%%%
\begin{adjustwidth}{-\extralength}{0cm}
	
	\reftitle{References}
	
	% Please provide either the correct journal abbreviation (e.g. according to the вЂњList of Title Word AbbreviationsвЂќ http://www.issn.org/services/online-services/access-to-the-ltwa/) or the full name of the journal.
	
	%\bibliography{References}
	
	%=====================================
	% References, variant B: internal bibliography
	%=====================================

\end{adjustwidth}

\end{document}